%% file: main.tex
\documentclass[a4paper]{article}

\usepackage{INTERSPEECH2022}
\usepackage{xcolor}
\usepackage{amsfonts,bbm}

\usepackage{caption}
\usepackage{tabularx}

\newcommand{\X}{\ensuremath{\mathcal{X}}}
\newcommand{\Z}{\ensuremath{\mathcal{Z}}}
\newcommand{\Y}{\ensuremath{\mathcal{Y}}}

\newcommand{\seqit}[1]{\bar{#1}}
\newcommand{\sy}{\seqit{y}}
\newcommand{\sz}{\seqit{z}}

\newcommand{\reals}{\mathbb{R}}

\usepackage{multirow}
\newcommand{\algoname}{\textit{DeepFry}}
\newcommand{\hubert}{\textit{HubertFry}}
\usepackage{url}
\newcommand{\secref}[1]{Section~\ref{#1}}
\makeatletter
\newcommand{\printfnsymbol}[1]{%
  \textsuperscript{\@fnsymbol{#1}}%
}
\makeatother

\title{DeepFry: Identifying Vocal Fry Using Deep Neural Networks}
\name{Bronya R. Chernyak\printfnsymbol{1}\thanks{\printfnsymbol{1}These authors contributed equally to this work. B.~R.~Chernyak is the corresponding author.}$^1$, Talia Ben Simon\printfnsymbol{1}$^2$, Yael Segal\printfnsymbol{1}$^1$, Jeremy Steffman$^3$, Eleanor Chodroff$^4$ , Jennifer S. Cole$^3$, Joseph Keshet$^1$}
\address{
$^1$Faculty of Electrical and Computer Engineering, Technion--Israel Institute of Technology, Israel\\
  $^2$Deptartment of Computer Science, Bar-Ilan University, Israel\\
  $^3$Deptartment of Linguistics, Northwestern University, USA\\
  $^4$Deptartment of Language and Linguistic Science, University of York, UK}
\email{chernroni@gmail.com}

\begin{document}

\maketitle
\begin{abstract}
Vocal fry or creaky voice refers to a voice quality characterized by irregular glottal opening and low pitch. It occurs in diverse languages and is prevalent in American English, where it is used not only to mark phrase finality, but also sociolinguistic factors and affect. Due to its irregular periodicity, creaky voice challenges automatic speech processing and recognition systems, particularly for languages where creak is frequently used.

This paper proposes a deep learning model to detect creaky voice in fluent speech. The model is composed of an encoder and a classifier trained together. The encoder takes the raw waveform and learns a representation using a convolutional neural network. The classifier is implemented as a multi-headed fully-connected network trained to detect creaky voice, voicing, and pitch, where the last two are used to refine creak prediction. The model is trained and tested on speech of American English speakers, annotated for creak by trained phoneticians.

We evaluated the performance of our system using two encoders: one is tailored for the task, and the other is based on a state-of-the-art unsupervised representation. Results suggest our best-performing system has improved recall and F1 scores compared to previous methods on unseen data.
\end{abstract}

\input{introduction}
\input{models}

\input{dataset}

\input{experiments}

\input{conclusions}

\bibliographystyle{IEEEtran}
\bibliography{mybib}

\end{document}

%% file: introduction.tex
\section{Introduction}\label{sec:intro}


Vocal fry, also known as Creaky Voice, is a type of phonation employed across a wide variety of languages with different linguistic and extralinguistic functions. A typical creaky voice has a low rate of vocal fold vibration (pitch), irregular pitch, and constricted glottis, characterized by a small peak glottal opening, long closed phase, and low glottal airflow \cite{keating2015acoustic}. Keating \emph{et al.} \cite{keating2015acoustic} outline several kinds of creaky voice, each of which manifests a slightly different set of acoustic properties. 

Cross-linguistically, creaky phonation plays many phonological roles \cite{davidson2021versatility}. It can serve as an utterance-final marker, signal phonemic contrasts with other voice qualities, or be an additional acoustic cue to enhance different contrasts, such as tone (as in Mandarin or Cantonese) \cite{yu2010laryngealization, kuang2017creaky}. It is used as a variant of glottal stop in many languages. Creaky phonation also plays a role in social interaction: it can indicate the end of a conversational turn (Finnish) \cite{ogden2001turn}, indicate an irritation (Vietnamese) \cite{mixdorff2003quantitative}, and be a marker to establish identities \cite{davidson2021versatility}. 


Creaky voice challenges automatic speech processing algorithms due to its irregular periodicity. As a result, algorithms for pitch tracking, spectral analysis, speaker verification, and automatic speech recognition might fail to operate in their full capacity \cite{cullen2013creaky, keshet2018automatic}. This is also a challenge in automatic processing of corpus phonetic and phonological analyses \cite{goldrick2016automatic, chodroff2017structure, chodroff2018corpus, hall2019phonological}. We believe the community will benefit from an open-source tool for automatic creak detection.



There have been several studies on algorithms for the detection of creaky voice. Early methods to detect creaky voice were based on ad-hoc signal processing techniques. Vishnubhotla and Espy-Wilson \cite{vishnubhotla2006automatic} proposed a set of rules on the AMDF measure of periodicity. Ishi \emph{et al.} \cite{ishi2007method} suggested to represent the speech by a pulse-synchronized analysis and then use a comparison of intra-frame periodicity and inter-pulse similarity. 

Kane, Drugman, Gobl \emph{et al.} \cite{cullen2013creaky, drugman2012resonator, scherer2013investigating, kane2013improved, drugman2014data} proposed several algorithms which are all based on specially designed acoustic features from the excitation and residual of the linear prediction filtering analysis of the speech and other acoustic features. This line of work used decision trees, fuzzy-input fuzzy-output support vector machine (F$^2$SVM) algorithm, and shallow artificial neural networks for the task. 
Recent work has proposed more advanced learning algorithms, but have been designed for a unique and small data set. Tavi {et al.} \cite{tavi2019recognition} suggested an exploratory creak recognizer based on a convolutional neural network (CNN), which is generated specifically for emergency calls. Villegas \emph{el al.} \cite{villegas2019prediction} used recurrent neural networks to detect creaky of single words in Burmese.

In contrast to previous work, we propose a deep learning model to detect creaky voice directly from an unprocessed speech signal. 
The model is built from an encoder and a multi-headed classifier trained together. The encoder is a CNN that takes the raw waveform and learns a representation of the signal. The classifier is a fully-connected network that gets as input the representation and outputs a detection score for creaky voice along with two additional auxiliary tasks: voicing and pitch. These additional predictions steer the overall network toward a better solution in detecting creaky voice \cite{shrem2019dr}.

We use two types of encoders. The first encoder is a specially designed encoder with a large time-span processing window proposed in \cite{segal2021pitch}. It has a larger receptive field than the standard receptive field used in speech processing, and is able to ``see'' several pitch periods, even for very low pitch values. Therefore, we believe that this encoder can contribute to the task of creaky voice detection. This is particularly useful compared to methods that operate on a processed windowed signal (such as MFCC, STFT, and others) and might lose pitch information crucial for identifying creak. 

The second encoder is based on a state-of-the-art self-supervised representation of the speech called HuBERT \cite{hsu2021hubert} that yields state-of-the-art results for downstream tasks, such as automatic speech recognition \cite{baevski2020wav2vec, hsu2021hubert2}. The latter encoder was pre-trained on 960 hours of read speech. The same multi-headed classifier was used in both models.

Trained phoneticians annotated two parallel corpora of connected speech each with 32 American English speakers (Datasets 1 and 2) along with a subset of speech from 14 American English speakers in the ALLSSTAR corpus \cite{bradlow2010allsstar}. The models were trained and developed on Dataset 1, and evaluated on Dataset 2 and the ALLSSTAR corpus. 

Results suggest that both models outperform a heuristics-based baseline model and have better F1 and recall values than Kane \emph{et al.} \cite{kane2013improved}. In addition, the model based on the pre-trained HuBERT representation was marginally better than the specially-designed encoder on unseen data. Our code and pre-trained models are publicly available here: {\footnotesize \url{https://github.com/bronichern/DeepFry}}.

%% file: models.tex
\section{Method}
\label{sec:models}

The input to all models is the raw waveform. Formally, we denote a speech waveform of $T$ samples by $\bar{x}=(x_1,\ldots,x_T)$, where $x_t\in\X$ for $t\in[1, T]$ and $\X\subseteq\reals$. Our setting is designed to allow different input duration hence $T$ is not fixed.  
We assume that there is a sequence of $K$ multi-labels, $\sy = (y_1,\ldots,y_K)$, where each multi-label $y_k$ is from the set $\Y = \{\text{(creaky, not-creaky), (voiced,unvoiced), (pitch,no-pitch)}\}$. 

\begin{figure}[h]
 \centering
\includegraphics[width=\linewidth]{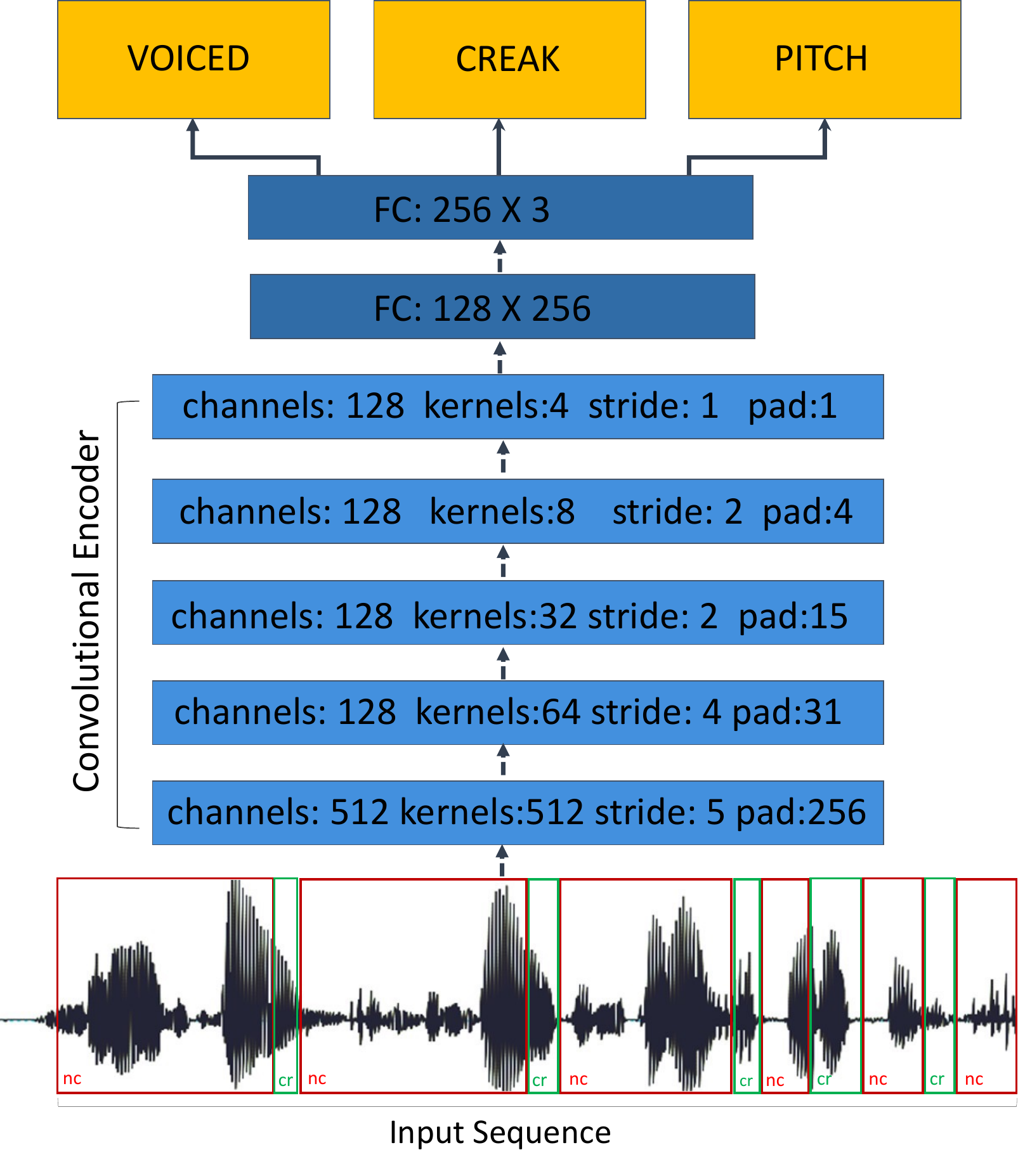}
 \vspace{-0.7cm}
 \caption{Model architecture. The encoder has 5 CNN layers, and the classifier has 2 fully-connected layers with 3 classification heads: creaky, voiced, pitch.}
 \label{fig:architecture}
\end{figure}

Our models are composed of two neural networks. The \emph{encoder} $g: \X\to \Z^K$ is a function from the domain of $\X$ to the embedding space $\Z \subseteq \reals^N$. Specifically, the encoder generates a sequence of representational vectors $\sz  = (z_1,\ldots,z_K)$, where $z_k \in \Z$, for all $1 \leq k \leq K$, such that $z_k\in\Z$ is the acoustic embedding of the $k$-th frame. The embeddings are then processed by a classifier that outputs a sequence of $K$ predictions. The \emph{classifier} is a function $f: \Z^K\to \Y^K$ from the domain of features vectors to the domain of target objects. 
Figure~\ref{fig:architecture} depicts an encoder and the multi-headed classifier. The encoder $g$ is a fully convolutional neural network and is based on the framework proposed in \cite{segal2021pitch} for pitch estimation. This encoder gets as input a variable duration signal and outputs a sequence of embedding vectors every 5 msec (due to hardware limitations, the number of embedding vectors was restricted to less than 100 per input). It has a larger receptive field than the standard receptive field used in speech processing, and therefore can handle several pitch periods. We denoted this encoder-classifier combination as \algoname. 


In our work we also use another encoder which is based on the HuBERT model \cite{hsu2021hubert}. This model learns a representation of the speech in a self-supervised manner. It is a large CNN model trained to distinguish a series of subsequent samples from random future samples when some representations are masked, similar to the BERT model \cite{devlin2019bert}. The rationale behind this concept is that subsequent samples are more likely to belong to the same phonetic class than random future samples. Here we used the HuBERT model that was pre-trained on 960 hours of read speech (LibriSpeech). The model operates on a 20 ms frame-rate window.  We denoted this encoder-classifier combination as \hubert. 

Our classifier is based on the concept of multi-task learning (MTL). MTL optimizes multiple tasks simultaneously, under the assumption that the information shared by they task will help boost the performance of the model on the task of interest.  Specifically our classifier $f$ is aimed at detecting creaky voice (binary),  voice/unvoiced (binary) and pitch (binary). The voiced/unvoiced was annotated by expert phoneticians, and the pitch was extracted using \cite{boersma1993accurate}.


%% file: dataset.tex
\section{Datasets}\label{dataset}
Three datasets of American English connected speech were annotated for creaky voice. 
The first two datasets investigated the relationship between information structure and the realization of nuclear and prenuclear pitch accents in American English.
The third dataset is the American English subset of the ALLSSTAR corpus (Archive of L1 and L2 Scripted and Spontaneous Transcripts and Recordings)\cite{bradlow2010allsstar}. Throughout the paper, the datasets are denoted as \textit{Nuclear}, \textit{Prenuclear} and \textit{ALLSSTAR}, respectively.

The \textit{Nuclear} and \textit{Prenuclear} datasets contain speech from 32 native speakers of American English (\textit{Nuclear}: 16 female, 16 male; \textit{Prenuclear}: 11 female, 21 male), and the \textit{ALLSSTAR} dataset contained speech from 14 native speakers of American English (7 female, 7 male).
All participants were university-aged students.

In the \textit{Nuclear} and \textit{Prenuclear} datasets, participants read aloud a series of three-sentence stories. 
Only the final `target' sentence was used in the present study of creak.
Each participant read all 20 target sentences (within a unique story) in a randomized order in four separate blocks. Only the first two blocks were analyzed. 
Block 1 was read in a neutral speaking style and block 2 in a lively speaking style. 
The \textit{ALLSSTAR} dataset contained readings of the first ten Hearing in Noise Test sentences in the ALLSSTAR HINT1 subset. 
The recordings in all three datasets were made at Northwestern University in a soundproof booth with a sampling rate of 22.05 kHz; all recordings were resampled to 16 kHz for the experiments. 
Sentences which contained a speech disfluency were removed from the analysis. As a result, 1195 sentences were available for analysis in the \textit{Nuclear} dataset, 1200 sentences in the \textit{Prenuclear} dataset, and 140 sentences in the \textit{ALLSSTAR} dataset . 


The \textit{Nuclear} dataset was divided into train (20 participants - about 24 minutes), validation (6 participants - about 6 minutes), and test (6 participants - about 8 minutes) folds where each fold contained an equal number of male and female participants. The \textit{Prenuclear} and \textit{ALLSSTAR} datasets were used for evaluation.

\subsection{Annotation}
To annotate the data, the transcripts were submitted to the Montreal Forced Aligner (MFA) for word- and phone-level alignments using the default American English pronunciation dictionary and acoustic model \cite{mcauliffe2017montreal}.
From these boundaries, we annotated the target sentences using two voice quality labels: modal or creaky voice, which are reasonably well-defined on sonorant segments.
An interval was labeled as modal voice if a visible pitch track was present in Praat over a sonorant region, and one of the following two scenarios held true: 1) the interval had audible modal voice quality (conveyed audible pitch, had a similar `smooth' voice quality as other prototypical modal regions), or 2) there was visible evidence of modal voicing from glottal pulses in the spectrogram or periodicity in the waveform.
An interval was labeled as creak if no visible pitch track was present over a sonorant region in Praat and the above criteria for modality were not met. 
The utterance boundaries were refined during annotation and resubmitted to the MFA for a more precise phone-level alignment. 
We note, that in our labeling process, only unvoiced phonemes could not be tagged as creak.


%% file: experiments.tex
\section{Experiments}
In this section, we present the results of our method. We begin by describing the hyper-parameters used to train our model. Then, we describe the measures used to evaluate our method and the adjustment we made to consider the different frame-rate of each method. Following that, we present ablation experiments of our model. We conclude this section by comparing our models to other methods on various unseen datasets and our test set.
We trained \algoname~for 14 epochs, with an Adam optimizer, a learning rate of 0.001, a dropout of 0.1 and batch size 16. 
\hubert~was trained for six epochs, with an Adam optimizer, a learning rate of 0.001, and a batch size of 16. Both models were trained on the \textit{Nuclear} dataset, and for both of them, we stipulated that the classification of a given frame was creak only if the frame was also predicted as voiced at inference time. 
\subsection{Evaluation method}
We evaluated all models using Precision (P), Recall (R) and F1 measures. Precision is defined as the number of creaky frames correctly classified by the algorithm divided by the total number of frames classified as creaky by the algorithm. Recall is defined as the number of creaky frames correctly classified by the algorithm divided by the total number of ground truth creaky frames. The F1 measure is defined as $F1 = {2PR}/{(P+R)}$.

Furthermore, our models, and the models we compare to, output predictions at a different frame-rate. Thus, for a fair comparison, all models classification was evaluated for a 20 ms frame-rate/window.

\subsection{Baseline model}\label{sec:baselines}

Our  baseline model (referred to as the `Praat baseline') used heuristics to detect the onset or offset of creaky voice for a given interval of speech \cite{boersma1993accurate}.
The heuristic operated on the assumption that modal voice could be detected by trackable numeric pitch values, and creaky voice would be associated with an undefined pitch value in Praat.
The script either moved \textit{start-to-end} or \textit{end-to-start} through Praat-extracted pitch values (time step = 0.15 s; pitch range = 50 to 350 Hz). 
When moving start-to-end, the script assumed modal voice was present until it encountered a sequence of undefined pitch values in frame $n$ and frame $n+2$.
When moving end-to-start, the script assumed creaky voice was present until it encountered a sequence of numeric pitch values in frame $n$ and frame $n-2$.
For intervals where the speech interval was not fully sonorant, creaky voice was frequently confused with a voiceless segment.


For the \textit{Nuclear} and \textit{Prenuclear} datasets, the start-to-end mode was used for the first three regions of the target sentence, namely the subject noun phrase, the verb and following determiner, and the object noun phrase. The final phrase and the full sentences in the \textit{ALLSSTAR} dataset were analyzed using the end-to-start mode. 

\begin{table}[h]
  \caption{The contribution of the classification heads for final predictions. Precision (P), recall (R), and F1 scores on the \textit{Nuclear} test set. Predictions are evaluated at a 20 ms frame-rate.}
  \label{tab:results_model1}
  \centering
  \begin{tabular}{@{\extracolsep{0.1pt}}lcccccc}
  \toprule
    \multirow{2}{*}{Models}& \multicolumn{3}{c}{Classification heads} &   \multirow{2}{*}{P} & \multirow{2}{*}{R} & \multirow{2}{*}{F1}   \\
    \cmidrule(r){2-4} 
    & Creak & \!\!\!Voice\!\!\! & Pitch & & \\
    \midrule
 \multirow{ 4}{*} \algoname&   $\surd$&&  &  \textbf{68.35}  &  69.64 &  68.99  \\
&   $\surd$&\!\!\!\!$\surd$\!\!\!\!& &  66.01 & 73.42 &  69.52 \\
   &   $\surd$&\!\!\!\!\!\!\!&$\surd$ & 67.30  &  69.67 & 68.46 \\
     &$\surd$&\!\!\!\!$\surd$\!\!\!\!&$\surd$& 67.43  &  \textbf{76.27}    &  \textbf{71.57}\\
\hline
\multirow{ 4}{*} \hubert &  $\surd$&&  & 68.65  &64.97  & 66.76  \\
  &   $\surd$ &\!\!\!\!$\surd$\!\!\!\!&&  \textbf{69.29} &64.52 &  66.82\\
    &   $\surd$&\!\!\!\!\!\!\!\!&$\surd$ & 68.07 & 68.66 & 68.36 \\
   &$\surd$&\!\!\!\!$\surd$\!\!\!\!&$\surd$& 68.74 &  \textbf{69.19}  & \textbf{68.96}   \\
     
    \bottomrule
  \end{tabular}
\end{table}

\begin{table*}[h]
  \centering
  \setlength{\tabcolsep}{1em}
\captionsetup{width=1.8\columnwidth}

\caption{Comparison to other methods. Precision (P), Recall (R), and F1 scores on different phonetic subsets of the test set of the \textit{Nuclear} dataset. Sonorants include vowels, glides, liquids, and nasals. Metrics are reported as percentages. Predictions are evaluated at a 20 ms frame-rate.}
\begin{tabular}{l ccc ccc ccc}
      \toprule
      \multirow{ 2}{*}{Models}    & \multicolumn{3}{c}{Vowels}  & \multicolumn{3}{c}{Sonorants} & \multicolumn{3}{c}{All}     \\
      \cmidrule(r){2-4}   \cmidrule(r){5-7}   \cmidrule(r){8-10}

       & P & R & F1      & P  & R  & F1     & P & R & F1     \\
    \midrule
    
Praat baseline &  59.66 &  59.21 & 59.43  &  60.07 & 60.60 & 60.33  & 32.93 &  60.60 & 42.67 \\
Kane \emph{et al.} \cite{kane2013improved} & \textbf{86.63}  & 30.14  & 44.72   &  \textbf{87.53}  &  28.53 &43.03  &  \textbf{75.98} &  28.53  &  41.48 \\

\algoname &  81.76   &   \textbf{75.71}  & \textbf{78.62} &  82.20  &   \textbf{76.27}  &  \textbf{ 79.12}  & 67.43  & \textbf{ 76.27}    &   \textbf{71.57}  \\
\hubert &  78.99    & 68.33 & 73.28&  79.69 &   69.19   & 74.07   & 68.74 &  69.19   & 68.96   \\
    \bottomrule
\end{tabular}
\label{tab:compare_nuc}
\end{table*}

\begin{table*}[h]

\captionsetup{width=1.8\columnwidth}
  \centering
    \caption{Evaluation on unseen data. Precision (P), recall (R), and F1 scores on different phonetic subsets of the \textit{Prenuclear} and \textit{ALLSSTAR} datasets. Sonorants include vowels, glides, liquids, and nasals. Metrics are reported as percentages. Predictions are evaluated at a 20 ms frame-rate.}
\begin{tabular}{llccccccccc}

    \toprule
       \multirow{ 2}{*}{Dataset}   &\multirow{ 2}{*}{Models}  & \multicolumn{3}{c}{Vowels} &  \multicolumn{3}{c}{Sonorants} & \multicolumn{3}{c}{All}    \\
        \cmidrule(r){3-5}   \cmidrule(r){6-8}   \cmidrule(r){9-11}
        &   &  P   & R   & F1      & P  & R  & F1     & P & R & F1    \\
    \midrule

   \multirow{ 4}{*}{\textit{Prenuclear}}&  Praat baseline\!\!& 44.01    &51.99 & 47.67 &  46.59    &  54.89   & 50.40 &  25.52    & 54.89  & 34.84\\
& Kane \emph{et al.} \cite{kane2013improved}\!\!&  \textbf{81.14}  & 47.58  & 59.99 &  \textbf{80.32} & 43.07 & 56.07      & 66.51 & 43.07   & 52.28 \\
& \algoname &    74.65   &     \textbf{77.50}      &    76.05      &    74.31        &   76.51      &    75.39    &     61.16      &   76.51     &  67.98     \\ 
& \hubert &    78.97   &  77.05 & \textbf{78.00} &  79.83 &   \textbf{78.24}  & \textbf{ 79.02}   & \textbf{69.53}  &  \textbf{78.24} &  \textbf{73.63}  \\
 \hline
    \multirow{ 4}{*}{\textit{ALLSSTAR}}&  Praat baseline\!\!&   85.34 &38.90 & 53.44 & 84.89 &  38.47  &52.95 & 26.60  &  38.47  &31.45\\
 & Kane \emph{et al.} \cite{kane2013improved}& \textbf{97.15}  &  36.50  & 53.06  &   \textbf{97.39} & 34.02 &  50.42 & \textbf{90.03} & 34.02  & 49.38\\
 & \algoname &   94.06     &   55.23      &  69.59       &    93.27        &   52.23      &       66.96 &    77.20       &   52.23     &   62.30    \\
&\hubert &   75.86   & \textbf{71.37} & \textbf{73.55} & 76.94  &    \textbf{70.33}  &  \textbf{73.49} & 66.74  & \textbf{70.33}     &    \textbf{68.49} \\
    \bottomrule
\end{tabular}

\label{tab:unseen}
\end{table*}
\subsection{Ablation}
We begin the experiments by analyzing the contribution of each additional task to the final detection of creak. To do so, we trained  \algoname~and \hubert~with the following settings: (i) only with creak head (ii) creak head and voice head (iii) creak head and pitch head (iv) with creak, voice, and pitch heads. 

Results are shown in Table \ref{tab:results_model1}, where each row represents a different experiment setting. Interestingly, it can be seen that each model benefits differently from each auxiliary task. 
By inspecting the F1 score, we can see that while \algoname~benefits more from the voice detection head than the pitch detection head, \hubert~presents higher performance with the pitch head. 
Finally, although training with each task individually does not necessarily significantly affect the performance, training with the combination of auxiliary tasks leads to increased performance.

\subsection{Comparison to previous works}
 

We now turn to compare our methods to previous works. We compared \algoname~and \hubert~to the baseline model presented in \secref{sec:baselines} on the test set of the \textit{Nuclear} dataset. We also compared to the model was proposed by Kane \emph{et al.} \cite{kane2013improved}\footnote{We use the open-source framework - Voice Analysis Toolkit \scriptsize\url{https://github.com/jckane/Voice_Analysis_Toolkit}}. We found no other implementations or a detailed enough description of the work mentioned in  \secref{sec:intro}.

Table \ref{tab:compare_nuc} presents the comparison for different subsets of phonemes, to gain a deeper understanding of the results. The first column, \textit{Vowels}, shows results only on the vowels phonemes. The second column, \textit{Sonorants}, contains the results on the subset which includes vowels, glides, liquids, and nasals phonemes. The last column, shows results on all the phonemes. Note that the results in column \textit{All} for \algoname~and \hubert~are the same results as in Table \ref{tab:results_model1}.


\algoname~received the best recall and F1 scores, while Kane \emph{et al.} \cite{kane2013improved} received the best precision scores.
Kane \emph{et al.}'s low recall suggests that the high precision is due to miss detection bias. In addition, Praat baseline outperformed Kane \emph{et al.} \cite{kane2013improved} on the \textit{Vowels} and \textit{Sonorants} subsets. Still, Praat baseline's precision score dramatically decreased when evaluated on all phonemes. That demonstrates this model's confusion between voiceless and creaky frames.

\subsubsection{Testing on unseen data}
Finally, we tested our models on the \textit{Prenuclear} and \textit{ALLSSTAR} datasets, which differ from the \textit{Nuclear} dataset we trained on. Both the \textit{Nuclear} and the \textit{Prenuclear} datasets have the same lexical content. However, the \textit{Prenuclear} and \textit{ALLSSTAR} datasets have a different set of speakers and phonetic realizations. Furthermore, the \textit{ALLSSTAR} dataset also has a different lexical content. 

Results are shown in Table \ref{tab:unseen}. It can be seen that in terms of F1 score, \hubert~outperformed the other methods across all datasets. However, Kane \emph{et al.}'s precision is the highest precision on the \textit{Vowels} and \textit{Sonorants}, which is consistent with the results reported in Table \ref{tab:compare_nuc}. Additionally, compared to Table \ref{tab:compare_nuc}, Praat baseline algorithm recall and F1 scores drop.
It is evident that \hubert~ achieved higher F1 and recall scores than \algoname~, still the results of \algoname~are relatively higher than the other methods we compared to. Nevertheless, we note that \hubert~was pre-trained on 960 hours, which is much more data than the 24 minutes of the \textit{Nuclear} dataset we trained \algoname~on. Furthermore, \hubert~has 95M parameters, compared to \algoname~which has less than 5M parameters. Therefore, we believe that training on more data, will improve the results of \algoname.

%% file: conclusions.tex
\section{Conclusions}
In this work, we present a system composed of an encoder and a classifier for creak detection. We investigated two types of encoders that work on the raw wave: (i) \algoname~ which is a small network with large receptive field and (ii) \hubert~which is a state-of-the-art encoder that was pre-trained on a lot of data and has a lot of parameters. Results suggest, that both of our implementations, achieve higher recall and F1 scores than the methods we compared to. Furthermore, it is evident that while \algoname~had better results on the test set of \textit{Nuclear} dataset, \hubert~had higher recall and F1 scores on datasets from different distribution. We believe this is due to the vast amount \hubert~was trained on. Thus, it remains for future work to investigate the effect of training with more data on \algoname.

\section{Acknowledgements}
Y. Segal is sponsored by the Ministry of Science \& Technology, Israel.